\documentstyle[12pt,moriond]{article}
\begin{document}

\heading{The Future of Millimetre and Submillimetre Cosmology }

\author{David H. Hughes}{Instituto Nacional de Astrofisica, Optica y Electronica
(INAOE), \\ Apartado Postal 51 y 216, 72000, Puebla, Pue., Mexico} {}

\begin{moriondabstract}

Using the submm array camera SCUBA on the 15-m JCMT it is now
possible to conduct unbiased submm surveys and quantify the level of
star-formation activity in the young Universe by observing the
rest-frame FIR thermal emission from dust in high-redshift
galaxies. However an accurate interpretation of these existing submm
cosmological surveys is prevented by uncertainties in the
redshifts of the submm-selected galaxies, the ambiguities in the
identifications of their optical/IR and radio counterparts and the
restricted range of flux density over which the submm source-counts
are measured. This paper outlines the future observations required to
overcome these deficiencies.

\end{moriondabstract}

\section{Introduction}
Observational evidence suggests that much of the on-going
star-formation in the young universe takes place in a heavily obscured
ISM and therefore must be {\em hidden} from extragalactic optical and
IR surveys.  Hence the {\em transparent} view of the high-$z$ universe
provided by submm--mm wavelength observations ($\lambda \sim 200 -
3000\mu$m) which are insensitive to the obscuring effects of dust, and
the strength of the negative k-correction which enhances the observed
submm--mm fluxes of starburst galaxies by factors of 3--10 at $z > 1$
\cite{Hughes99}, offer obvious advantages.

The opportunity to conduct cosmological observations at submm--mm
wavelengths has been realised in the last few years with the
successful development and commissioning of sensitive bolometer arrays
({\it e.g.}  SCUBA-I, SHARC-I, BoloCam-I, MPIfR 37-channel).  These
arrays, which will be upgraded within 1--2 years, operate on the
largest submm and mm telescopes (15-m JCMT, 10-m CSO, 30-m IRAM).

\section{Submillimetre Cosmological Surveys}

By early 2001 the first series of extragalactic SCUBA (850~$\mu$m)
surveys will be completed, covering areas of 0.002--0.12~deg$^2$ with
respective $3\sigma$ depths in the range $\rm 1.5~mJy < S_{\nu} <
8~mJy$ \cite{Smail97}, \cite{Hughes98}, \cite{Barger98},
\cite{Eales99}, \cite{Lilly99}.  The evolution of the high-$z$
starburst galaxy population can be determined from an accurate measure
of the integral submm source-counts, the FIR luminosities, star
formation rates (SFRs) and redshift distribution of the submm selected
galaxies. The contribution of the submm sources to the total FIR - mm
background places an additional constraint on the competing models.
To ensure these submm data are fully exploited, the current SCUBA
surveys have been restricted to fields that have been extensively
studied at other wavelengths (X-ray, optical, IR and radio) and have
yielded the following preliminary results:

\begin{itemize}

\item
The faint submm source-counts at 850$\mu$m are reasonably well
determined between 1--10\,mJy and significantly exceed a no-evolution
model, requiring roughly $(1+z)^{3}$ luminosity evolution out to
$z\sim 2$, but with poor constraints at higher redshifts
(fig.\,1). The submm background measured by COBE requires that the
SCUBA source-counts must converge at $\rm S_{850\mu m} \leq
0.5$mJy. Approximately 30--50\% of the submm background has been
resolved into individual galaxies with flux densities $ S_{850\mu
\rm{m}} > 2\,$~mJy.

\item 
Submm sources generally appear to be associated with $z > 1$ galaxies,
although it not clear whether they necessarily have optical, IR and
radio counterparts.  There is still much debate about the fraction of
submm sources at $z\geq 2$, and the fraction of submm sources that
contain an AGN.

\item
At high-redshift ($2 < z < 4$) the sub-mm surveys appear to find
$\sim5$ times the star formation rate observed in the optical surveys,
although the effects of dust obscuration and incompleteness in the
optical are still uncertain.
\end{itemize}

\subsection{Limitations on an understanding of high-$z$ galaxy evolution}

Despite the success of the first SCUBA surveys, we can 
identify the deficiencies in the submm data which prevent a more
accurate understanding of the star-formation history of high-$z$
galaxies.  This paper summarises these deficiencies and outlines the
future observations which will alleviate the following problems.\\

\noindent{\underline{\em {\rm 2.1.1} Poorly constrained evolutionary models}}
\hspace{2mm}  To improve the constraints on the competing evolutionary models
provided by the current submm source-counts, it is necessary to (1)
extend the restricted wavelength range of the surveys, (2) increase
the range of the flux densities over which accurate
source-counts are measured and (3) increase the number of sources
detected at a given flux level by surveying greater areas.  

Ground-based surveys at mm wavelengths can take advantage of a more
stable and transparent atmosphere which will provide increased
available integration time (to gain deeper survey sensitivity or
greater survey area) and increased flux calibration accuracy.  Future
surveys with more sensitive and larger format arrays ({\it e.g.}
BoloCam) operating at 200$\mu$m -- 3\,mm on airborne and ground-based 
telescopes will allow signifcantly greater areas to be covered (hence
more sources detected) and will increase the range of the flux densities
over which sources are detected (fig\,1.).  The deepest surveys todate
are still only sensitive to high-$z$ galaxies with SFRs comparable to
the most luminous local ULIRGs ($\geq 200 M_{\odot} \rm
yr^{-1}$). Furthermore conducting surveys with larger diameter
telescopes ({\it e.g.} 50-m Gran Telescopio Milimetrico (LMT), 100-m
GBT) will reduce the beam-size, hence decrease the depth of the
confusion limit (allowing deeper surveys) and improve the positional
accuracy of detected sources.\\

\begin{figure*}[t]
\vspace{6.5cm}
\includegraphics{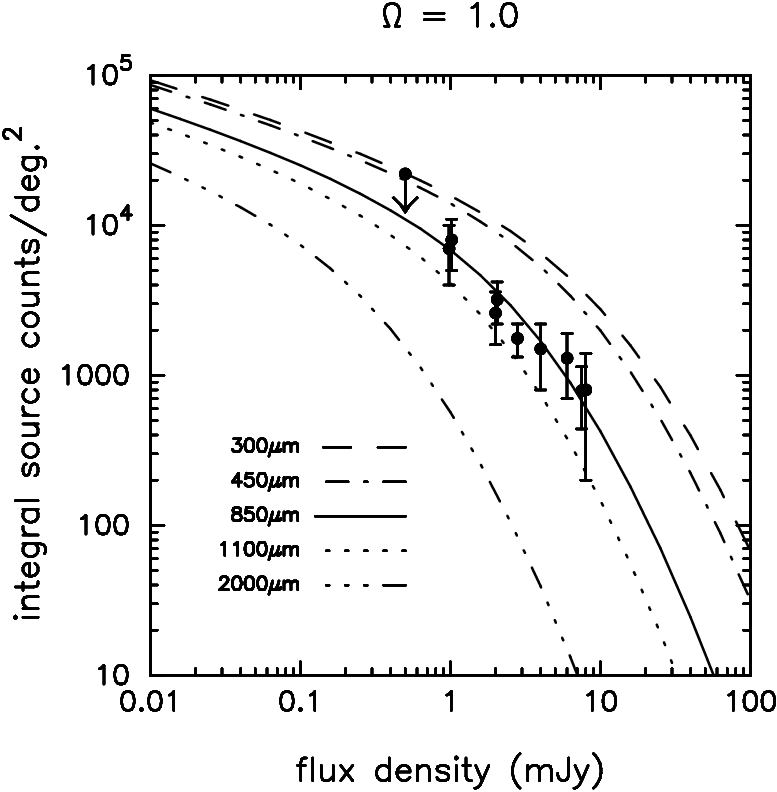}
\includegraphics{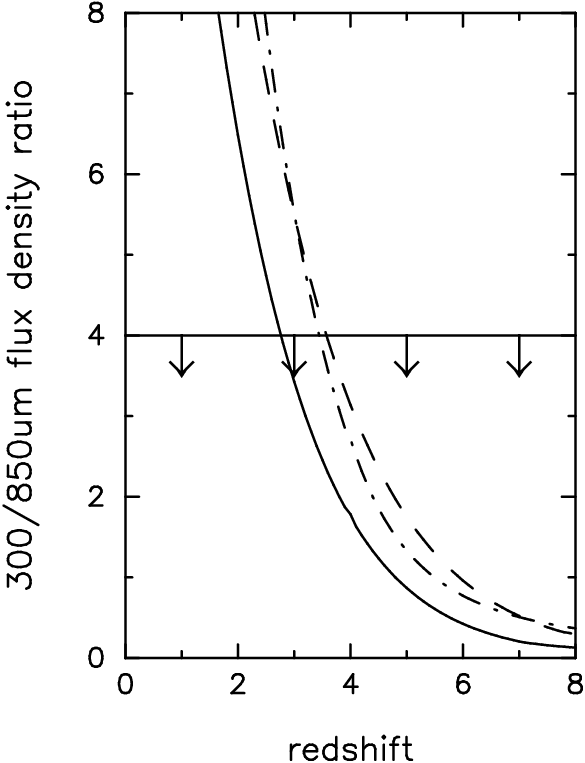}
\caption{\footnotesize Predicted number-counts at submm and mm
wavelengths.  The data represent the measured 850$\mu$m source-counts
from the SCUBA surveys described in \S\,2. A representative
evolutionary model at 850$\mu$m (pure luminosity evolution of the form
$(1+z)^{3.0}$ out to $z_{u}$=2.3, with evolution then held constant
between $2.3 \leq z_{c} < 6.0$) is extrapolated to derive the number
counts at 300--2000$\mu$m. The data are consistent with a range of
models (where $z_{u} \sim 2.0 - 3.0$ with an upper-bound of $z_{c} >
5.0$). More accurate submm and mm source-counts at brighter ($>$10\,mJy)
and fainter ($<$1\,mJy) flux densities are required to fully constrain
the competing models.}

\caption{\footnotesize The 300/850$\mu$m flux ratio, appropriate for the
combination of BLAST and SCUBA surveys (\S\,2.1.2), is a powerful
discriminant of redshift. The example of a 4$\sigma$ 850$\mu$m
detection (10~mJy), from the medium-depth UK SCUBA survey, with no
BLAST 300$\mu$m counterpart ($< 40$\,mJy) is indicated by the
horizontal line. This upper limit to the 300/850$\mu$m ratio implies a
redshift $> 3$ assuming the SED of the high-$z$ source is similar to
the range of low-$z$ starburst and AGN SEDs represented by the solid
(Arp220), dashed (M82) and dashed-dotted (Mkn231) curves.}
\end{figure*}

\noindent {\underline{\em {\rm 2.1.2} Ambiguity in the optical counterparts \&
redshifts of submm galaxies}} \hspace{2mm} The current SCUBA surveys
(with $15''$ resolution at 850$\mu$m) are struggling to unambiguously
identify the submm sources with their optical/IR/radio
counterparts. Hence the redshift distribution and 
luminosities of the submm sources are still uncertain.  This results
directly from the submm positional errors of $\sim 2-3^{\prime\prime}$
that are typical for even the highest S/N submm detections, and from
the lack of submm data measuring the redshifted FIR
spectral peak at 200--450~$\mu$m.

The positions of the brightest SCUBA sources ($S_{850\mu m} > 8$\,mJy)
can be improved with mm-interferometric observations. However our IRAM
Plateau de Bure follow-up of the brightest source in the Hubble Deep
Field has demonstrated that even with $\leq 2''$ resolution
and sub-arcsec positional errors, ambiguous optical identifications,
and hence ambiguous redshifts remain \cite{Downes99}.  It
should be no surprise that submm selected galaxies, including those
with mm-interferometric detections, do not always have optical
counterparts since high-$z$ galaxies observed in the earliest stages of
formation may be heavily obscured by dust. Indeed this is the most
compelling reason for conducting the submm surveys in the first
instance and therfore searches for the counterparts may be
more successful at near-infrared wavelengths. This was recently
demonstrated by Smail {\it et al.} \cite{Smail99} who took deep near-IR
($2\mu$m) images of two lensed clusters previously observed by SCUBA
\cite{Smail97}.  The original counterparts were identified
as bright low-redshift ($z \sim 0.4$) galaxies 5--10 arcsecs
distant from the submm sources.  However the new IR images revealed
two high-$z$ ($z > 2$) IR galaxies, with no optical counterparts,
within 2--3 arcsecs of the SCUBA sources.  The obvious consequence of
these misidentifications is an inaccurate determination of the 
star-formation history of high-$z$ starburst galaxies.  

The uncertainty in the redshift distribution of submm-selected
galaxies can be significantly reduced by measuring the mid-IR to radio
SEDs of the individual sources.  The power of using mid-IR to radio
flux ratios ({\it e.g.} 15/850$\mu$m, 450/850$\mu$m,
850$\mu$m/1.4~GHz) as a crude measure of redshift 
was demonstrated by Hughes {\it et al.} \cite{Hughes98}
during the SCUBA survey of the Hubble Deep Field and has since been
described elsewhere \cite{Carilli99}, \cite{Blain99}.  
Given sufficient sensitivity, the mid-IR--submm--radio colours of a submm
source can discriminate between optical/IR counterparts which are
equally probable on positional grounds alone, but which have
significantly different redshifts, $\delta z \geq 2$, (fig.\,2).
This important technique, and the necessity for sensitive short submm
data (200--500$\mu$m) measuring the rest-frame FIR SEDs of the
individual high-$z$ submm galaxies, without which it remains impossible
to constrain their bolometric luminosities and SFRs,
provide the major scientific justifications behind {\em BLAST}, a possible
future long-duration Balloon-borne Large Aperture Submm Telescope
(P.I. M.Devlin, University of Pennsylvania).

\begin{table}[t]
\begin{center}
\begin{tabular}{ccccccc}
\hline \hline
survey area         & 
1$\sigma$ depth     & 
no. of pixels       & 
\multicolumn{2}{c}{no. of detected galaxies}   & 
\multicolumn{2}{c}{no. of $>5\sigma$ galaxies} \\ 
(sq. degrees) & 
              & 
              & 
$> 5  \sigma$  & 
$> 10 \sigma$  & 
$ z > 1$   & 
$ z > 3$   \\
\hline \hline
\multicolumn{7}{c}{6 hour 300~$\mu$m test-flight survey: D=2.1~m,
$\theta = 41''$, 
NEFD=150 mJy s$^{1/2}$} \\  \hline
0.24 & 7  mJy & 2352  & 120 & 34 & 110 & 18 \\
0.55 & 10 mJy & 4800  & 150 & 40 & 135 & 20 \\
1.1  & 15 mJy & 10800 & 135 & 30 & 125 & 16 \\
\hline
\multicolumn{7}{c}{50 hour 300~$\mu$m long-duration flight survey: D=2.7~m,
$\theta = 32''$, 
NEFD=90 mJy s$^{1/2}$} \\  \hline
3.3  &  7 mJy & 32884  & 1670 & 480 & 1530 & 250 \\
6.8  & 10 mJy & 67111  & 1870 & 500 & 1680 & 250 \\
15.4 & 15 mJy & 151000 & 1890 & 420 & 1740 & 220 \\ \hline
\end{tabular}
\caption{Number and $z$-distribution of galaxies detected in 
alternative BLAST surveys.}
\end{center}
\end{table}

Alternative {\em BLAST} 300$\mu$m surveys based on the model shown in
fig\,1 are described in Table\,1.  Assuming a 3$\sigma$ 300$\mu$m
confusion limit of $\sim 20-30$\,mJy, a single 6\,hour {\it BLAST}
test-flight survey can follow-up the widest of the current SCUBA
surveys, detecting all $>\,4\sigma$ SCUBA sources ($\sim 100$ sources
with $S_{\rm 850\mu m} >\,10$\,mJy) in a 0.24 sq. deg. survey at all
redshifts $< 3$.  Non-detections at 300$\mu$m imply $z > 3$.
Increasing the primary aperture of {\em BLAST} to 2.7-m and conducting
a 50-hour survey during a long-duration balloon flight significantly
increases the survey area and number of sources detected to $> 1500$,
a comparable number to that detected by BoloCam in a future 50-hour
0.45 sq. degree 1100$\mu$m survey with the GTM/LMT.

An accurate determination of the redshift distribution of submm
selected galaxies will ultimately be achieved through the measurement
of mm-wavelength $^{12}$CO spectral-line redshifts, without recourse
to having first identified the correct optical or IR counterparts.
This ``CO {\em redshift machine}'' requires a large instantaneous
bandwidth ($\Delta\nu \sim 30-40$\,GHz) to take advantage of the
reduced separation of adjacent mm-wavelength $^{12}$CO transitions in
the high-$z$ universe, $\delta \nu_{J, J-1} \sim
115/(1+z)$\,GHz. Hence at redshifts $ > 2$ any adjacent pair of
$^{12}$CO lines are separated by $< 40$\,GHz, the frequency separation
defining the precise redshift of the galaxy.  The combination of data
from BLAST, JCMT, CSO and the GTM/LMT will efficiently pre-select from
submm surveys, using their FIR--mm colours, those galaxies with
sufficiently high (but still unknown) redshifts that are suitable
targets to follow-up with a ``CO {\em redshift machine}''.


%
\vfill
\end{document}